# A Rapid and Robust Automated Macroscopic Wood Identification System using Smartphone with Macro-lens

Xin Jie Tang*, Yong Haur Tay*, Nordahlia Abdullah Siam**, Seng Choon Lim**

Center of Computing and Intelligent Systems(CCIS),

Universiti Tunku Abdul Rahman*,

Forest Research Institute of Malaysia**

Abstract

Wood Identification has never been more important to serve the purpose of global forest species protection and timber regulation. Macroscopic level wood identification practiced by wood anatomists can identify wood up to genus level. This is sufficient to serve as a frontline identification to fight against illegal wood logging and timber trade for law enforcement authority. However, frontline enforcement official may lack of the accuracy and confidence of a well trained wood anatomist. Hence, computer assisted method such as machine vision methods are developed to do rapid field identification for law enforcement official. In this paper, we proposed a rapid and robust macroscopic wood identification system using machine vision method with off-the-shelf smartphone and retrofitted macro-lens. Our system is cost effective, easily accessible, fast and scalable at the same time provides human-level accuracy on identification. Camera-enabled smartphone with Internet connectivity coupled with a macro-lens provides a simple and effective digital acquisition of macroscopic wood images which are essential for macroscopic wood identification. The images are immediately streamed to a cloud server via Internet connection for identification which are done within seconds.

Introduction

Illegal logging is one the main cause of deforestation contributing to global warming and adverse environmental issues like changes in natural habitat for animal in the forest. Moreover, unregulated logging drives protected wood species to near extinction. USA Lacey Act, the European timber regulation (EUTR), Illegal Logging Prohibition Act in Australia are laws enacted to protect forests and to regulate timber trading (Dormontt, 2015).

Wood identification provides one of the most valuable supports in combatting illegal logging. The methods for the macroscopic and microscopic wood identification are well established and have been used for more than 100 years. More recent development on wood identification includes computer-aided machine vision identification systems based on visual and textual descriptions (Koch, 2015).

Wood Identification has never been more important to serve the purpose of global forest species protection and timber regulation. Macroscopic level wood identification practiced by wood anatomists can identify wood up to genus level (Gasson, 2011). This is sufficient to serve as a frontline identification to fight against illegal logging and timber trade for law enforcement authority. This is the first but critical step in law enforcement in combating illegal logging because it can prevent logging and forest can be better protected. United Nations Office on Drugs and Crime published a practical guideline on Best Practice Guide for Forensic Timber Identification in 2016 (Anon. 2016) shows the importance of wood identification.

However, there are huge number of timber types across different regions which may take time in building talents for wood identification. Furthermore, it is costly to deploy expert on wood identification in the frontline to combat illegal logging. These challenges threatens the effort of combating illegal logging and timber regulation as frontline identification is one of the effective counter measures. Therefore, it is not uncommon for one to raise questions on the consistency, accuracy and accessibility of the frontline enforcement. How consistent is that the enforcement official within a region? How accurate are they among the agency?

Hence, computer assisted method such as machine vision methods are developed to do rapid field identification for law enforcement official. Machine vision methods aim to replicate a wood expert on wood identification. A general overview of a machine vision system is detailed in Hermanson *et al.*, (2011). Tou *et al.*, (2007) and Khalid *et al.*, (2008) used the machine vision system in labarotory setting that can identify wood species. More recently, XyloTron by Hermanson *et al.*, (2013) from USDA Forest Products Laboratory provides a field deployable wood identification system that comprises of a customized camera with a computer running machine vision algorithm program. These systems and approaches prove that computer-aided machine vision system is viable in macroscopic wood anatomy identification. A common characteristics of Khalid *et al.*, (2008) system and XyloTron system is a specialized camera module with structured light source. This is to ensure the consistency of the image acquisition process to capture well-lit macroscopic wood image so that the machine vision model can identify the timber types correctly. However, this setup brought some drawbacks to the scalability of the application.

Firstly, Khalid *et al.*, (2008) camera module was not meant to be portable as it was designed for laboratory. XyloTron's camera module solved the portability problem by custom fabricating a camera module with imaging sensors, controlled illumination light source and some connection ports to communicate with an external computer running the machine vision program. This setup is field-deploy ready but it involves some works to deploy it to frontline like the assembly of the camera module and installation of software.

In this paper, we proposed a rapid and robust macroscopic wood identification system using machine vision method with off-the-shelf smartphone and retrofitted macro-lens. Our system is cost effective, easily accessible, fast and scalable at the same time provides human-

Rapid and Robust Macroscopic Wood Identification System 5

level accuracy on identification. Camera-enabled smartphone with Internet connectivity coupled with a macro-lens provides a simple and effective digital acquisition of macroscopic wood images which are essential for macroscopic wood identification. The images are immediately streamed to a cloud server via Internet connection for identification which are done within seconds.

A general overview of our system (Fig. 1) is to replicate human wood anatomist on identifying macroscopic wood images. The core of the system is a machine learning model that learns to identify macroscopic wood images. The following sections depict the architecture of the system.

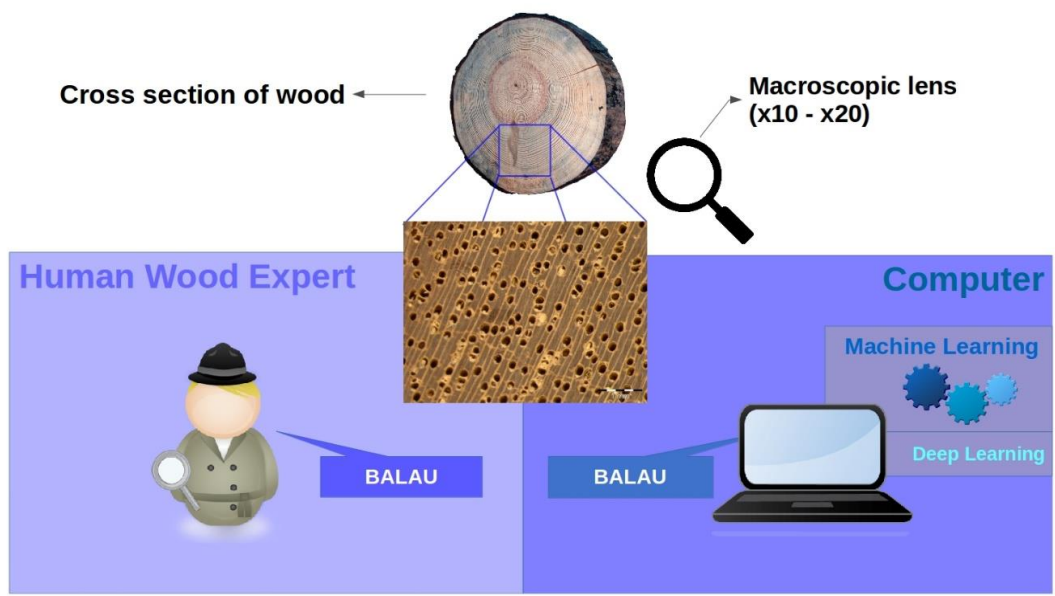

*Fig. 1*: A machine learning model replicating a human anatomist on identifying macroscopic wood images based on the macroscopic wood anatomy.

Architecture

The main idea of our system is a trained artificial intelligence model with machine learning capability to match human level accuracy which utilizes the speed and computing power of computer for wood identification. Machine learning algorithm is used to train on professionally labelled and verified macroscopic wood samples to produce a reliable, consistent and robust human level identification system. There are two major components for the system: A trained machine learning model and image acquisition method as shown in Figure 2.

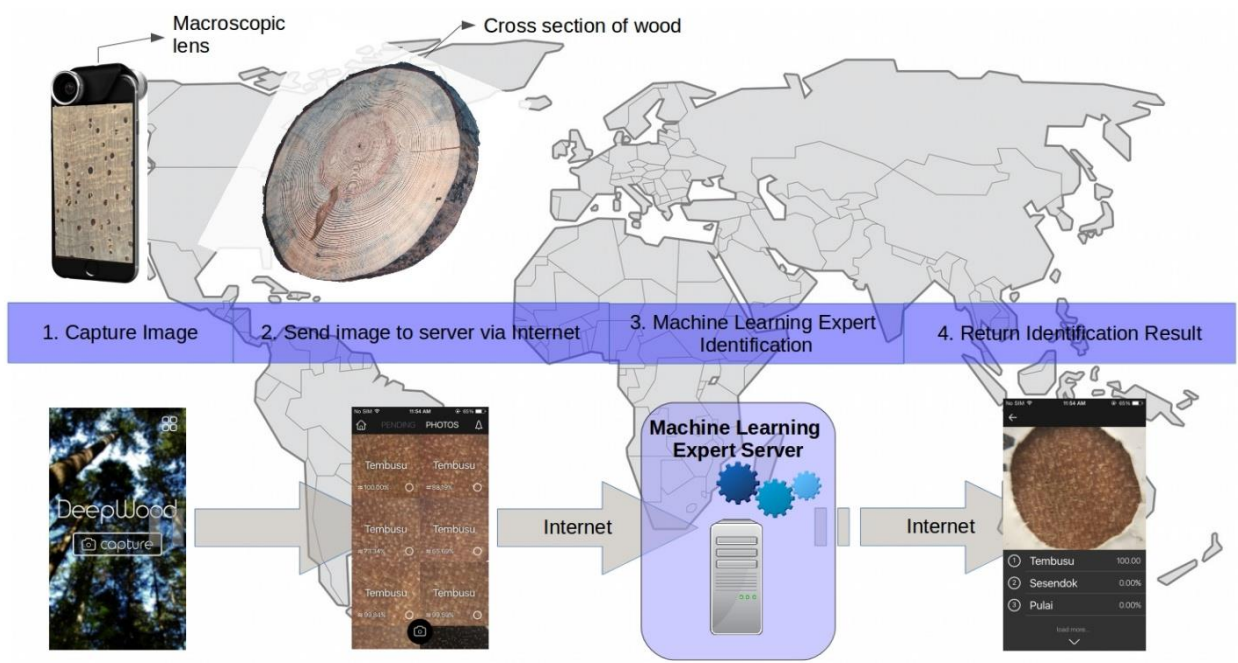

*Fig. 2*: General overview of the system architecture which consists of a smartphone with fitted macro-lens along with installed application, prepared wood samples, Internet connectivity and cloud based machine learning wood identification system.

Image Acquisition Device:

Smartphones are no longer a luxurious product and they have been changing our way of life in almost any level of connectivity and has become a common necessity. With Internet connectivity, smartphone has connected us to the world with unprecedented level of closeness. Smartphones are multifunctional with imaging capability and Internet connectivity, and its small form-factor allows great mobility. Hence, it enjoys great mass adoption which inspires us to use it as an image acquisition device as it offers great accessibility that no any other specialized imaging device can offer.

Besides mobility, smartphone cameras technology has come a long way to a point where they are comparable to digital single-lens reflex (DSLR) camera quality where average smartphone can capture images that are stunning with great details. Due to the ever increasing demand in smartphone photography, customized retrofitted lens are developed to cater the enthusiasm of smartphone photographer. Hence, macroscopic lens that offers magnification from the range of 5x to 30x can be easily bought and retrofitted by users to capture macroscopic images.

Our proposed smartphone with macroscopic lens setup as imaging device solves two challenges; lower initial cost and faster field deployment time. However, the trade-off of this off-the-shelf setup is that we lose the illumination control on images. On field identification, this is difficult to ensure consistency on the quality and illumination of the images and hence may jeopardize the performance of the identification system. To mitigate the problem of inconsistency of image quality, machine learning algorithm is used to utilize the power of computer to learn from real world dataset. In short, our imaging setup fulfills these requirements: great accessibility, cost effective as compared to specialized imaging devices, great mobility, and great connectivity.

Machine Learning Process:

The core of the system is a machine learning model that learns from lots of real world macroscopic wood samples. Deep learning, an active research area in machine learning grabs the spotlight in Artificial Intelligence recently, gives computer or machine the ability to learn from and make prediction on data. Deep learning, a deep neural network architecture has achieved some amazing performance in ImageNet 1000 class problem image classification (Deng *et al.*, 2009).

Generally, there are mainly two types of Deep Learning or Machine Learning namely supervised learning and unsupervised learning. Supervised learning is like a teacher teaching a student. For our application, the learning material is a huge dataset of professionally labelled macroscopic wood images. The students are then trained to identify timbre types based on the dataset. This mimics human learning experience from going through lots of learning material in order to master a knowledge. Using the raw computational power of computer, learning to identify timber types takes hours as opposed to weeks or months for human.  Besides that,  machine is emotionless and consistent.

The capability of Deep Learning algorithm inspires us to apply it to solve the inconsistency of captured wood image by providing a huge amount of wood images with huge variation to train the Deep Learning model. The next question, what kind of data should the machine learns in order to identify timber types?

Learning Material:

Using back the teacher and student analogy, the performance of a student greatly depends on the difficulty and the coverage of the learning material. If a student is trained on simple task, he or she might not be able to perform well in a difficult situation. Hence, in order to trained a practical Deep Learning expert, we try to mimic the standard practices of human anatomist on identifying macroscopic timber types.

In real world practices, a simple knife cut on the tangential cross section of woods exposes unique wood anatomy features as depicted in Koch (2015) and Best Practice Guide for Forensic Timber Identification United Nations Office on Drugs and Crime (Anon. 2016). This simple cut is to expose the structure of wood clearly. After that, a hand lens is used to study and examine the wood structure to perform identification. This is where our proposed system comes in, to replicate a Wood Anatomist with machine learning algorithm to identify timber types based on the macroscopic structure.

To ensure practicality, the learning material should be as close as possible to real world field macroscopic wood images. Hence, a rigorous and relentless data collection has been carried out to collect real world macroscopic wood images on wood samples that are prepared with standard procedure as mentioned above using our proposed smartphone camera with macro-lens setup. As a result, a massive dataset consists of 80,000 images of 60 timber types were collected, labelled and verified by Wood Anatomists. With the learning material prepared, it is ready to train the machine to become a human anatomist. Some of the collected data is shown in Figure 3 below.

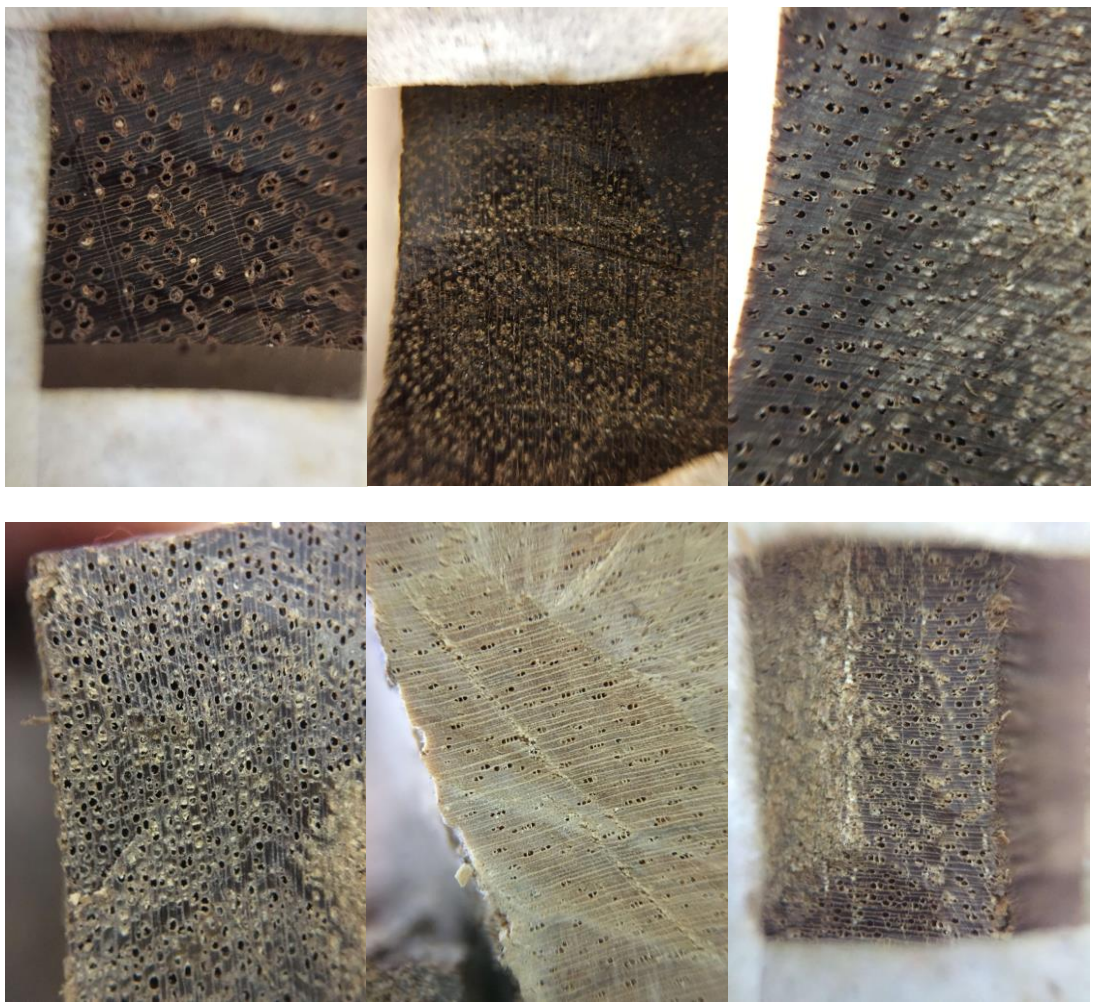

*Fig. 3*: Real world collected data on macroscopic wood images. From top left: Merbau, Chengal, Dark Red Meranti, Keruing, Jelutong, and Balau. Wood samples were prepared with a simple knife cut to expose wood structure and captured with a smartphone camera with macro-lens.

Cloud Computing:

With a core Deep Learning expert ready, we faced another problem which is how to deploy it to the frontline. Utilizing the global connectivity of Internet, cloud computing allows application to be reached by end users from around the world. Instead of running application on local personal computer or local server, application server hosted in the cloud provides easy access, zero maintenance and hassle-free user experience.

The core system is maintained by dedicated software engineers as opposed to local running program that requires some manual works by user on installation. Our system applies a service based model where users just have to install the application from mobile online store such as AppStore by Apple and can directly access the service.

This architecture allows the core system to be constantly updated and the images captured can be shared across for verification purposes by the off-site human experts, if need arises.

End User Perspective:

Our system provides a solution to rapid field identification on virtually any part of the globe that has Internet coverage. Taken into account of limited or slow Internet access, our system is also designed to enable offline image capturing. This feature allows user to capture wood images on wood in questioned without the connecting to the Internet. When the smartphone is connected to Internet, previous captured unidentified images will be streamed to the application server for identification. The overall process usually takes 3 to 4 seconds depends on the quality of Internet connection.

The image capturing to getting the result is shown in Fig. 2. From the user's perspective, however, there is a need to load the application in the smartphone, fix the macro-lens, prepare the surface of the wood to expose a clear wood structure and finally capturing the wood image using the smartphone. The process is now completed after the server returns an identification result. The flow is intuitive, simple and fast. With simple training on the proper way of cutting the wood, any user from different background or expertise can perform a rapid field identification on wood easily.

Prototype Application Findings

Through the collaborative research by Forest Research Institute of Malaysia (FRIM) and Universiti Tunku Abdul Rahman (UTAR), Malaysia, a prototype application was developed with some interesting findings. In this application, the Deep Learning wood expert is trained on 60 types of tropical timber types as summarized in Table 1. An average of 30 samples for each timber types were used to train the Deep Learning model.

This application features a simple camera interface for capturing image, a gallery storing captured wood images and predictions, and a wood information page. The application is designed to follow the flow as depicted in Figure 2. Capture, wait for result, and goto wood information page for more detailed descriptions on a particular timber type.

The identification system achieved a Top-1 accuracy of 96% and a Top-2 accuracy of 98%. Note that Top-2 accuracy signifies that the real wood type is in within the highest 2 predictions by the system. Upon some analysis, we found that the system is confused within the Top-2 predictions among timber types from same family, e.g. Dipterocarpaceae. For instance, a Dark Red Meranti is identified as Light Red Meranti. This is due to the similarity on wood structure found on the same family timber types.

## Conclusion

We presented a simple off-the-shelf smartphone camera with macro-lens setup as image acquisition device to capture macroscopic wood images for identification. Utilizing the learning capability of Deep Learning algorithm, we mitigate the loss of consistent illumination as provided by specialized camera module as per XyloTron setup. Cloud hosted identification systems allows us to deploy an artificial wood expert with human level performance on the frontline wherever is covered with Internet connection.

Tables

Table 1

*List of 60 Tropical Timbers for prototype application*

| | | |
|---|---|---|
| Ramin | Merpauh | Bintangor |
| Karas | Machang | Geronggang |
| Acacia mangium | Terentang | Medang |
| Meranti Bakau | Rengas | Belian |
| Mersawa | Merbau | Keledang |
| Balau | Keranji | Terap |
| Chengal | Kempas | Jelutong |
| Keruing | Kekatong | Tembusu |
| Dark Red Meranti | Tualang | Taek |
| Light Red Meranti | Sepetir | Penarahan |
| White Meranti | Sena | Pulai |
| Yellow Meranti | Kungkur | Kulim |
| Gerutu | Mengkulang | Durian |
| Melantai | Kembang Semangkok | Mempisang |
| Kasai | Nyatoh | Melunak |
| Resak | Bitis | Kelat |
| Kapur | Mata Ulat | Punah |
| Merawan | Perupok | Simpoh |
| Red Balau | Rubberwood | Kedondong |
| Giam | Sesendok | Seraya White |